\documentclass[twocolumn,pre,superscriptaddress,floatfix,longbibliography]{revtex4-1}

\usepackage{graphics}
\usepackage{graphicx}
\usepackage{amsmath}
\usepackage{amssymb,xcolor}
\usepackage{gensymb}
\usepackage{ulem}

\newcommand{\textin}[1]{\mbox{\scriptsize{#1}}}

\definecolor{grisclair}{rgb}{0.6,0.6,0.6}

\newcommand{\beq}{\begin{equation}}
\newcommand{\ee}{\end{equation}}

\begin{document}

\title{Influence of surfactant kinetics on rapid interface creation via microjet impact on liquid pools}
\author{D. Fern\'andez-Mart\'{\i}nez}
\address{Depto.\ de Ingenier\'{\i}a Mec\'anica, Energ\'etica y de los Materiales and\\ 
Instituto de Computaci\'on Cient\'{\i}fica Avanzada (ICCAEx),\\
Universidad de Extremadura, E-06006 Badajoz, Spain}
\author{E. J. Vega}
\address{Depto.\ de Ingenier\'{\i}a Mec\'anica, Energ\'etica y de los Materiales and\\ 
Instituto de Computaci\'on Cient\'{\i}fica Avanzada (ICCAEx),\\
Universidad de Extremadura, E-06006 Badajoz, Spain}

\author{J. M. Montanero}
\address{Depto.\ de Ingenier\'{\i}a Mec\'anica, Energ\'etica y de los Materiales and\\ 
Instituto de Computaci\'on Cient\'{\i}fica Avanzada (ICCAEx),\\
Universidad de Extremadura, E-06006 Badajoz, Spain}

\author{U.J. Guti\'errez-Hern\'andez}
\address{Mesoscale Chemical Systems Group, MESA+ Institute and Faculty of Science and Technology, University of Twente, PO Box 217, 7500 AE Enschede, The Netherlands}
\address{Departamento de F\'isica, Facultad de Ciencias, Universidad Nacional Aut\'onoma de México, Circuito Exterior S/N, Ciudad Universitaria, Ciudad de M\'exico 04510, M\'exico.}

\author{D. Fern\'andez Rivas}
\address{Mesoscale Chemical Systems Group, MESA+ Institute and Faculty of Science and Technology, University of Twente, PO Box 217, 7500 AE Enschede, The Netherlands}

\begin{abstract}
We experimentally investigate the influence of surfactant adsorption kinetics on cavity dynamics during the rapid formation of interfaces. For this purpose, we use a submillimeter jet impacting onto a surfactant-laden liquid pool much larger than the jet dimensions. Cavity retraction and closure occur on a submillisecond timescale, posing a stringent test of the ability of surfactants to reduce surface tension dynamically. Our experiments reveal the difference between the effects of sodium dodecyl sulfate (SDS), a surfactant with moderately fast adsorption kinetics, and Surfynol 465, a surfactant with ultrafast adsorption kinetics. For SDS, the collapse pathway is nearly indistinguishable from that of pure water, suggesting negligible dynamic surface tension reduction. In contrast, Surfynol allows the emergence of deeper cavities that persist longer in the liquid pool. The harmonic oscillator model accurately captures the cavity retraction in the deep seal regime. The fitted values of the damping ratios are consistent with the dynamic surface tensions.
\end{abstract}

\maketitle

\section{Introduction}

The formation and evolution of liquid–fluid interfaces occur in a wide range of physical, biological, and technological processes. The presence of surfactants heavily influences the dynamics of these interfaces. Surfactants dissolved in a liquid phase adsorb at the liquid-fluid interface and alter interfacial mechanics \citep{A16,MS20,NED21,PPS22,VM24,M24}. Surfactants are nearly unavoidable in real environments and can regulate many interfacial processes even at trace concentrations \citep{RCVMC25}. At higher concentrations, surfactant can drastically change multiphase flows on the millimeter and submillimeter scales. Some relevant examples include droplet coalescence \citep{OJ19,CBKSCJM21} and fragmentation \citep{KWTB18,PRHEM20}, bubble stability \citep{RVCMLH24,HLFCM25}, air entrainment \citep{CKBSCJM21,PKSCJM24,VM24}, and tip streaming \citep{A16,M24}, among many others.

Surfactant adsorption kinetics at liquid–air interfaces are recognized as a key factor shaping transient interfacial properties during rapid flows \citep{VSQCC24, MS20, QCSSCC25, CKSCJMCC23}. When a new surface is created (e.g., during jet or drop impact on a solid surface or a liquid pool), the initial surface tension practically equals that of the pure solvent. Then, the surface tension progressively decreases as surfactant molecules diffuse and adsorb to the new interface. The timescale over which dynamic surface tension approaches its equilibrium value is set by either diffusion of the surfactant molecule, the interfacial adsorption energy barrier, or a combination of both \citep{MS20}. In many cases, this timescale is much larger than that characterizing the surface growth. Only ``ultrafast"\ surfactants (those with very high adsorption rates) significantly reduce the surface tension during the rapid interface creation in phenomena such as the jet or drop impact, which takes place on timescales of the order of a few milliseconds. Standard surfactants fail to act on those timescales, even though they produce a considerable reduction in surface tension at equilibrium.  

The effect of surfactant kinetics has been observed in droplets impacting onto both a solid surface \citep{VSQCC24} and surfactant-laden pools and liquid films \citep{QCSSCC25, CKSCJMCC23}. \citet{VSQCC24} showed that only ultrafast surfactants significantly affect the maximum spreading diameter of a droplet impacting a smooth substrate. These surfactants trigger splashing or rim breakup at much lower Weber numbers than those of standard surfactants when the droplet impacts onto a surfactant-laden pool. Both numerical and experimental studies have demonstrated that Marangoni stress can delay or reshape the dynamics of interface retraction, particularly when the surfactant efficiently populates the rapidly expanding interface.

A new interface is rapidly generated when a train of droplets or a jet strikes a liquid pool. Researchers have mapped out the universal stages of this ballistic process: expansion, retraction, and closure via interface pinch-off of the cavity formed by the impact \citep{BHCFOLSM16, SPBT18}. This cavity can develop features similar to those seen in rigid-body impacts \citep{AB09}. A balance of inertia and both capillary and viscous forces governs these stages. Three main cavity closure pathways have been distinguished: shallow seal, deep seal, and surface seal. The selection of the closure pathways can be predicted from the values of the Weber $\text{We}=\rho v_j^2 r_j/\sigma$ and Ohnesorge $\text{Oh}=\mu/(\rho r_j \sigma)^{1/2}$ numbers characterizing the problem (here $v_j$ and $r_j$ are the jet velocity and radius, respectively, $\rho$ and $\mu$ are the density and viscosity of the liquid pool, respectively, and $\sigma$ the liquid-air surface tension). 

\begin{figure}[!]
\begin{center}
\resizebox{0.95\columnwidth}{!}{\includegraphics{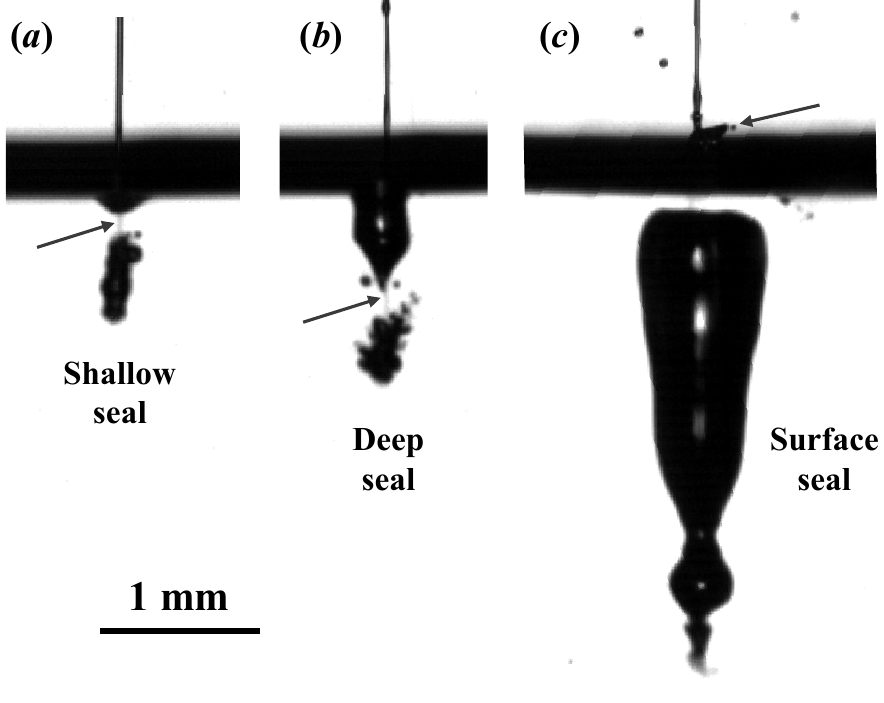}}
\end{center}
\caption{Cavity closure modes: shallow seal (a), deep seal (b), and surface seal (c). The images were taken with pure water for a jet radius $r_j=$ 30.9 $\mu$m (a), 36.4 $\mu$m (b), and 38.5 $\mu$m (c) and velocity $v_j=$ 14.9 m/s (a), 19.9 m/s (b), and 29.3 m/s (c), respectively. The arrows indicate the pinch-off location.}
\label{closure}
\end{figure}

Consider a submillimeter jet impacting on a low-viscosity liquid pool (characterized by a low Ohnesorge number). The shallow seal occurs for low enough Weber numbers. In this case, the maximum penetration depth of the cavity is relatively small, and the interface pinches close to the pool's free surface (Fig.\ \ref{closure}a). Higher Weber numbers result in the deep seal closure, where the cavity typically penetrates deeper into the pool and the interface pinches farther from the surface (Fig.\ \ref{closure}b). In these regimes, the cavity may experience successive pinch-offs at multiple positions. Lastly, the surface seal occurs for sufficiently high Weber numbers. In this case, the surface closure occurs while the cavity is still expanding, and the interface pinches above the pool's surface (Fig.\ \ref{closure}c). Similar cavity profiles have recently been reported when the target object is a droplet or a capillary liquid bridge \citep{QHMF21, QF23}. In all of these cases, high-speed jets and droplets create cavities that expand and collapse within submillisecond timescales. Numerical simulations reveal how the transition between capillary-driven deep seal closure and gas-pressure-driven surface seal closure becomes sensitive to impact parameters and fluid properties, providing a new methodology on rapid interfacial dynamics \citep{KFQ24}.

Directly linking surfactant kinetics to measurable mechanical outcomes during high-speed interface creation remains a challenge. To address this problem, we utilize submillimeter jet impact on liquid pools. This phenomenon allows us to create a new interface at a controlled rate that can match or exceed the adsorption rate of the surfactant dissolved in the liquid pool. Our objective is to report new jet impact experiments and to utilize them as a diagnostic for surfactant kinetics.

In this study, we compare the effects of Sodium Dodecyl Sulfate (SDS) and Surfynol 465 \citep{PZPA16} on the impact of submillimeter jets on a water pool. These surfactants produce practically the same decrease in the surface tension of the air-water interface at equilibrium. However, the adsorption rate of Surfynol is supposed to be significantly larger than that of SDS, as suggested by the experiments of \citet{VSQCC24}. Our experimental configuration enables us to select a hydrodynamic timescale that is much smaller than that characterizing SDS adsorption but comparable to that of Surfynol, thereby revealing the role played by surfactant kinetics. We observe significant differences between the penetration depths and cavity retraction times of the experiments with SDS and Surfynol. These experiments show how the surfactant kinetics set limits on reducing surface tension during rapid events. Our findings may provide a valuable reference for the selection of surfactants when rapid interface formation occurs in technological or medical applications, such as liquid microjet impingement for electronic cooling \citep{WZL25}, high-speed emulsification systems \citep{KRMCAS18, ZL22}, or in needle-free injection systems \citep{SF22, VMQF23}, where enhanced spread and adsorption of the drug in the target tissue is desired.

The manuscript is organized as follows. Section \ref{formulation} formulates the problem and provides the governing dimensionless parameters. Section \ref{methods} details our experimental method. Section \ref{results} presents a comparison of cavity evolution for the two surfactants. Section \ref{conclusions} summarizes the main findings of the work.

\section{Formulation of the problem}
\label{formulation}

\begin{figure}[!]
\begin{center}
\resizebox{0.6\columnwidth}{!}{\includegraphics{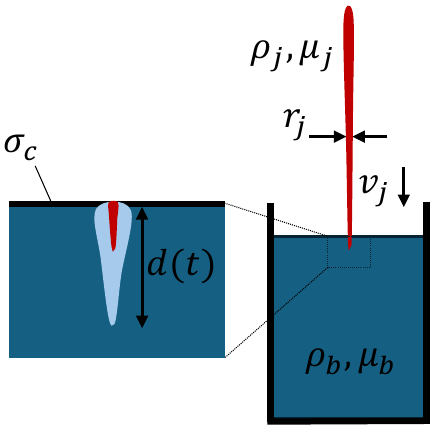}}
\end{center}
\caption{Sketch of the fluid configuration. A submillimeter jet (red) impacts on a centimeter liquid pool (blue), creating a cavity filled with air (cyan).}
\label{Probl_Sketch}
\end{figure}

Consider a liquid bath (pool) with density $\rho_b$ and viscosity $\mu_b$. A liquid jet of density $\rho_j$, viscosity $\mu_j$, radius $r_j$, and velocity $v_j$ hits the bath free surface (Fig.\ \ref{Probl_Sketch}). The surface tension of the air-liquid bath interface in the absence of surfactant is $\sigma_c$. We will analyze the evolution of the cavity resulting from the jet impact. Specifically, we will determine the cavity penetration depth $d$ as a function of time $t$ ($t=0$ corresponds to the jet impact instant).

The initial surfactant concentration on the liquid pool surface is $\Gamma_0$ (the surface coverage, measured in mols per unit area). The surfactant strength can be quantified by the maximum packing concentration $\Gamma_{\infty}$. In addition, ${\cal D}$ and ${\cal D}_s$ are the volumetric and surface diffusion coefficients, respectively. In the Langmuir model, the adsorption and desorption constants $k_a$ and $k_d$ characterize the surfactant sorption kinetics \citep{MS20}.

It is instructive to formulate the problem in dimensionless terms. Assuming that the gas phase is air and in the absence of surfactant, the dimensionless depth $\hat{d}=d/r_j$ depends on the density and viscosity ratios, $\rho=\rho_j/\rho_b$ and $\mu=\mu_j/\mu_b$, the Ohnesorge number Oh$=\mu_b/(\rho_b r_j \sigma_c)^{1/2}$, the Weber number $\text{We}=\rho_b v_j^2 r_j/\sigma_c$, and the dimensionless time $\hat{\tau}=t/t_c$, where $t_c=r_j/v_j$ is the characteristic time. The dimensions of the pool are assumed to be much larger than the jet radius and the cavity penetration depth. This ensures that boundary effects are negligible throughout the jet impact and the resulting cavity evolution. Therefore, the problem does not involve any geometrical parameters other than $r_j$. The Bond number $B=\rho_b g r_j^2/\sigma_c$ takes values of the order of $10^{-3}$, and, therefore, gravity can be neglected.

In the presence of an insoluble surfactant, the parameters characterizing the surfactant effect are the initial (equilibrium) surface coverage $\hat{\Gamma}_0=\Gamma_0/\Gamma_{\infty}$, the Marangoni number Ma=$\Gamma_{\infty} R_g T/\sigma_c$, where $R_g$ and $T$ are the gas constant and liquid temperature, respectively, and the surface Peclet number $\text{Pe}_s=r_j v_c/{\cal D}_s$ number, where $v_c=\sigma_c/\mu_b$ is the viscous-capillary velocity. In the presence of solubility effects, one must also consider the bulk Peclet number $\text{Pe}=r_j v_c/{\cal D}$. The surfactant solubility is characterized by the dimensionless adsorption and desorption constants, $\hat{k}_a=\Gamma_{\infty} k_a/v_c$ and $\hat{k}_d=r_jk_d/v_c$. We conclude that
\begin{equation} 
\hat{d}=\hat{d}(\rho,\mu,\text{Oh},\text{We};\hat{\Gamma}_0,\text{Ma},\text{Pe}_s;\hat{k}_a,\hat{k}_d,\text{Pe};\hat{\tau}).
\end{equation}

We will compare the impact of jets on a water pool containing either SDS or Surfynol. In both cases, the jets are made of the same liquid and have the same radius and velocity. The surfactants do not significantly alter the liquid pool density and viscosity. Consequently, the values of $\rho$, $\mu$, $\text{Oh}$, and $\text{We} $ are the same in the two experiments. The surfactant concentrations are such that $\Gamma_0\simeq \Gamma_{\infty}$. Therefore, $\hat{\Gamma}_0\simeq 1$ in all the experiments with surfactants. The maximum packing concentration $\Gamma_{\infty}$ (and, therefore, Ma) takes similar values in the two surfactants. The Surfynol diffusion coefficients are unknown. However, it can be expected to be similar to those of SDS because the molecular weights of the two molecules are comparable (the molecular weights of SDS and Surfynols are 288 g/mol and 666 g/mmol, respectively). Therefore, we can assume that the Peclet numbers also take similar values in the two experiments. The jet impact entails the rapid creation of a new interface to which surfactant adsorbs. This implies that $\hat{k}_a$ is expected to be more important than $\hat{k}_d$ in our experiments. We conclude that the difference between $\hat{d}=\hat{d}(\hat{\tau})$ in the experiments with SDS and Surfynol must be attributed essentially to the adsorption coefficient $\hat{k}_a$. We will change the characteristic time of interface production by changing the jet velocity (the Weber number). This will enable us to assess the impact of the adsorption rate on the dynamics of the cavity. Table \ref{t2ñ} shows the values of some of the above-mentioned dimensionless numbers.

\begin{table}
    \begin{tabular}{cccccccc}
    \hline
        & $\rho$ & $\mu$ & \text{Oh} & $\Gamma_0$ & Ma & $\text{Pe}_s$ & $\text{Pe}$ \\
     \hline
    SDS & 1.0 & 0.91 & 0.021 & 1.0 & 0.11 & $1.3 \times 10^3$ & $1.3 \times 10^3$  \\
    \hline
    Surfynol 465 & 1.0 & 0.91 & 0.021 & 1.0 & 0.071 & -- & -- \\
    \hline
    \end{tabular}
\caption{Values of the dimensionless numbers involving the physical properties of the fluids and surfactant monolayer.}
\label{t2ñ}
\end{table}

\begin{figure}[!]
\begin{center}
\resizebox{0.95\linewidth}{!}{\includegraphics{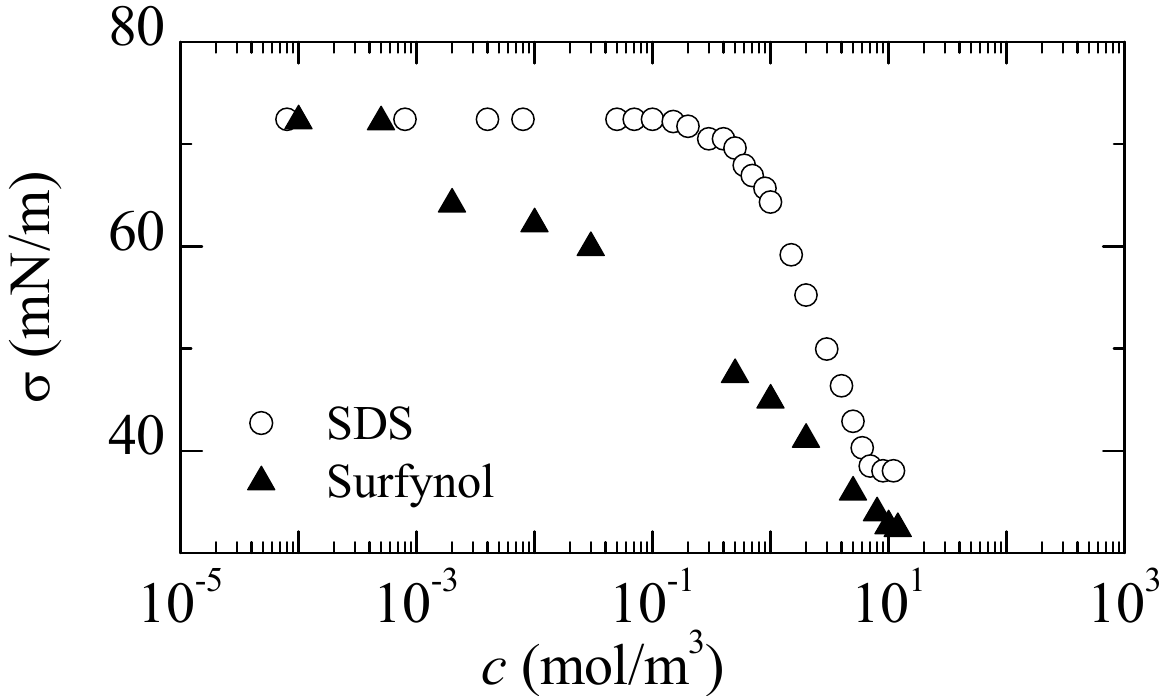}}
\end{center}
\caption{Surface tension $\sigma$ as a function of the surfactant volumetric concentration $c$ for SDS and Surfynol.}
\label{st}
\end{figure}

We will consider dimensional quantities in our analysis to show the spatial and temporal scales of the problem. It is worth noting that the characteristic length $r_j$ and time $t_c$ take practically the same values in each pair of SDS and Surfynol experiments. Therefore, the discussion in Sec.\ \ref{results} on the differences between $d(t)$ obtained with the two surfactants can be extrapolated to the differences between their dimensionless counterparts $\hat{d}(\hat{\tau})$. As mentioned above, these differences can essentially be attributed to the values of the adsorption constant for SDS and Surfynol, with the adsorption constant for Surfynol being unknown.

\section{Experimental Method}
\label{methods}

\begin{figure*}[!]
\begin{center}
\resizebox{0.8\linewidth}{!}{\includegraphics{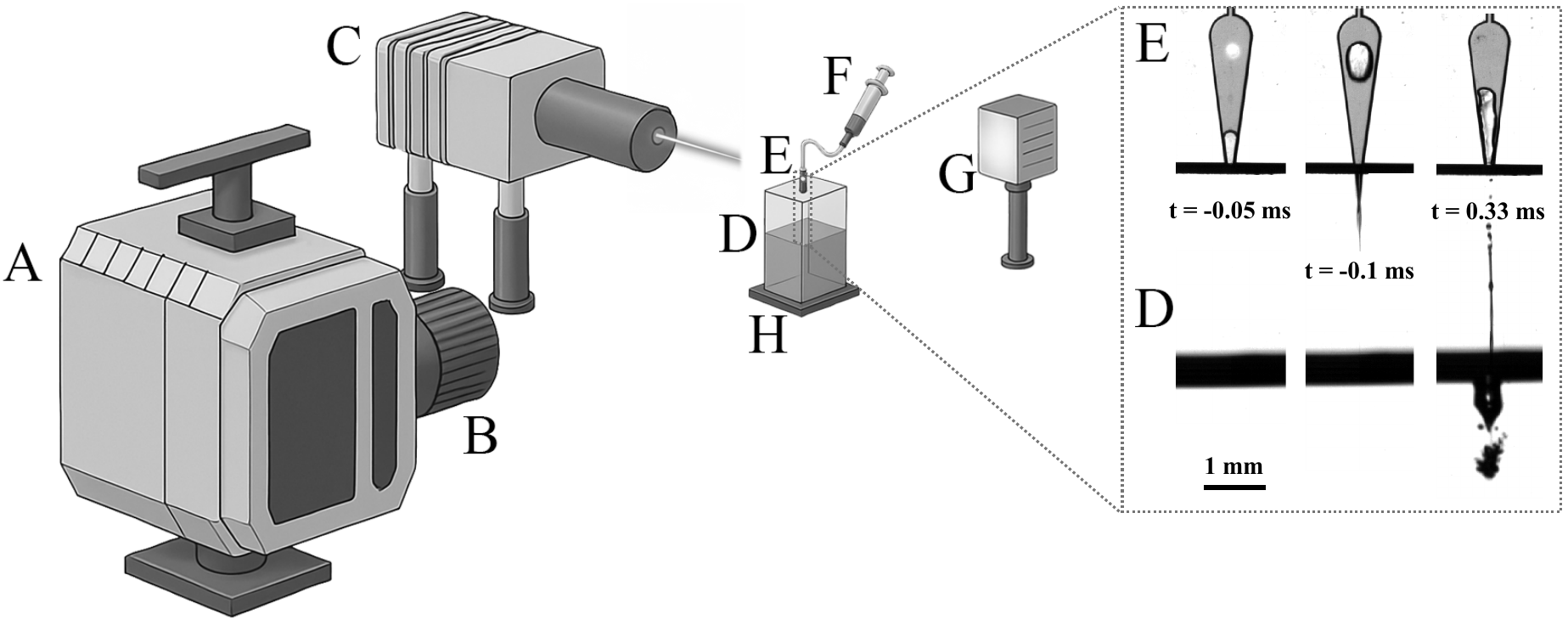}}
\end{center}
\caption{Schematic representation of the experimental setup: (A) high-speed camera, (B) optical lenses, (C) laser, (D) liquid bath, (E) microfluidic chip, (F) syringe, (G) light source, (H) translation stage. The inset shows a typical snapshot sequence of the thermocavitation and jet generation process.}
\label{Exp_Sketch}
\end{figure*}

Our experiments follow the protocol established in Refs.\ \citep{QF23, KFQ24} to produce submillimeter jets from the laser-induced cavitation. A custom-fabricated borosilicate glass microfluidic chip was filled with an aqueous solution of Allura Red AC (Sigma Aldrich) at 0.5 wt\% to enhance laser absorption. Jet generation was achieved by tightly focusing a pulsed laser (Litron, Nano S, 4 ns, 532 nm) at the chip base. To ensure reproducibility of the jet size and velocity, the bubble location and filling level in the liquid channel were kept constant. The induced pressure pulse and the rapid bubble expansion drive the formation of a high-speed submillimeter jet. The jet radius, measured far from the nozzle, was approximately 35 $\mu$m, and the jet velocity varied between 9 m/s and 33 m/s, corresponding to Weber numbers in the range of 34 to 527. These values were obtained by carefully tuning the laser energy between 172.3 $\mu$J and 580.6 $\mu$J. The Supplemental Material shows illustrative videos of the jet target.

The receiving pool walls were made of acrylic and mounted beneath the jet axis. The dimensions of the pool were $24\times 24\times 48$ mm$^3$, larger than $10^3$ times the jet radius and the cavity penetration depth. We conducted experiments with ultra-pure Milli-Q water and aqueous solutions of SDS (Sigma-Aldrich) at 8.1 mM and Surfynol 465 (Evonik) at $10$ mM. These concentrations approximately correspond to the critical micelle concentrations. Figure \ref{st} shows the surface tension $\sigma$ as a function of the surfactant volumetric concentration $c$ for SDS and Surfynol. Experiments were conducted at $21 \pm 1^\circ$C.

The bubble nucleation, jet formation, jet impact, and cavity evolution were captured using a high-speed camera (Photron Fastcam SAX2) coupled to a $12\times$ Ultra Zoom Navitar lens set at $1\times$ magnification level. The images were taken at 40000 fps with an exposure time of 10 $\mu$s over a typical experiment duration of 20 ms. The images were analyzed with a custom-made MATLAB script to obtain the evolution of the cavity profiles. Figure \ref{Exp_Sketch} shows the experimental setup. 

\section{Results and Discussion}
\label{results}

\subsection{Cavity dynamics and types of closure}

\begin{figure*}[!]
\begin{center}
\resizebox{0.8\textwidth}{!}{\includegraphics{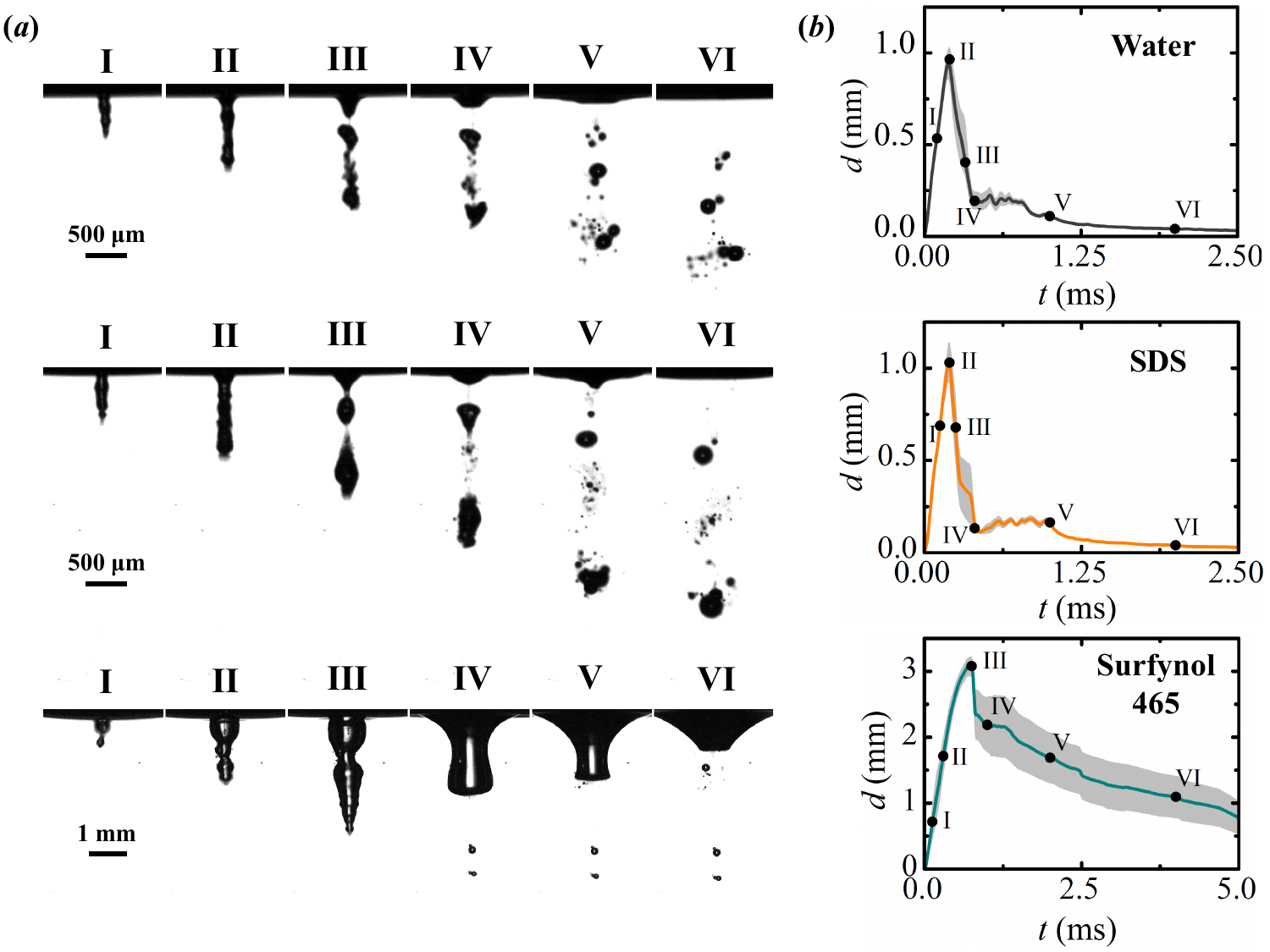}}
\end{center}
\caption{(a) Images of the cavity evolution for $r_j\approx$ 39 $\mu$m, $v_j \approx 20$ m/s, and the three liquids analyzed in this work (top: water, middle: SDS, bottom: Surfynol). The images correspond to the instants indicated in the graphs. (b) Mean cavity penetration depth $d(t)$. The shaded gray area represents the uncertainty in the estimate of the mean.}
\label{Illustration}
\end{figure*}

Figure \ref{Illustration} shows representative images of the cavity evolution obtained for $r_j\approx 39$ $\mu$m and $v_j\approx 20$ m/s. The sequence of images for SDS is similar to that of pure water. In the two cases, the cavity closure can be categorized as a shallow/deep seal, characterized by fast interface recoil driven by capillarity. This suggests that the properties of the cavity-pool interface remain essentially unchanged, as the adsorption rate of SDS is insufficient to fill the new interface created by cavity penetration. Our results align with recent studies \citep{VSQCC24, QCSSCC25}, which found that slow or moderately fast surfactants exert little influence on interfacial dynamics because interface creation outpaces surface renewal timescales. 

It is worth considering the difference between the cavity growth in this problem and the well-known ejection of Worthington jets by bubble bursting \citep{VM24}. The initial stage of cavity growth in our problem substantially differs from that leading to the Worthington jet emission in bubble bursting. In our problem, we hypothesize that the jet's impact on the pool surface creates a surfactant-free region that serves as the cavity precursor. The cavity interface grows from that clean region. Therefore, surfactant molecules cannot be dragged to the cavity apex. Conversely, the Worthington jet in bubble bursting arises from a surfactant-enriched region of the bubble \citep{VM24}. Surface convection drags the surfactant molecules and accumulates them in the jet tip. This results in a Marangoni stress directed to the jet base, which partially immobilizes the interface. 

The sequence of images for Surfynol differs significantly from that of water and the SDS aqueous solution (note that the scale bar in the images of Surfynol is different). The cavity initiated in the Surfynol-laden pool exhibits a pronounced growth phase, followed by a slow, symmetric retraction. The process leans toward a persistent deep seal: the cavity penetrates deeper before slowly retracting, resulting in a more drawn-out evolution. This effect is consistent with trends observed in drop-on-liquid interface experiments \citep{CM17,QCSSCC25}. The effect of Surfynol can be attributed to its rapid adsorption at the newly created surface, which lowers the surface tension. 

One may wonder whether the differences in cavity growth with SDS and Surfynol can be explained in terms of Marangoni stress caused by the Surfynol concentration gradient. However, Marangoni stress is typically associated with the partial immobilization of the interface \citep{MS20}. This is contrary to what we observe in our experiments, where the interface is mobilized (expands faster than in the clean water) by adding Surfynol.  

The differences between the cavity evolution are quantified in the graphs of Fig.\ \ref{Illustration}b, which show the cavity penetration depth $d(t)$. The maximum penetration depth for water is practically the same as for the SDS aqueous solution, while it increases by about a factor of three in the presence of Surfynol. The retraction is much slower in the presence of Surfynol. We will examine this last process in more detail in Sec.\ \ref{sec43}.

We conclude that noticeable differences in cavity dynamics appear depending on the kinetics of the surfactant dissolved in the liquid pool. When the pool contains an ultrafast surfactant, the cavity grows deeper and retracts considerably slower than in pure water or an aqueous solution of a standard surfactant. In other words, in the presence of an ultrafast surfactant, the cavity reaches a greater depth and persists longer before retracting and closing. 

\subsection{Maximum cavity depth before pinch-off}

\begin{figure*}[!]
\begin{center}
\resizebox{1\textwidth}{!}{\includegraphics{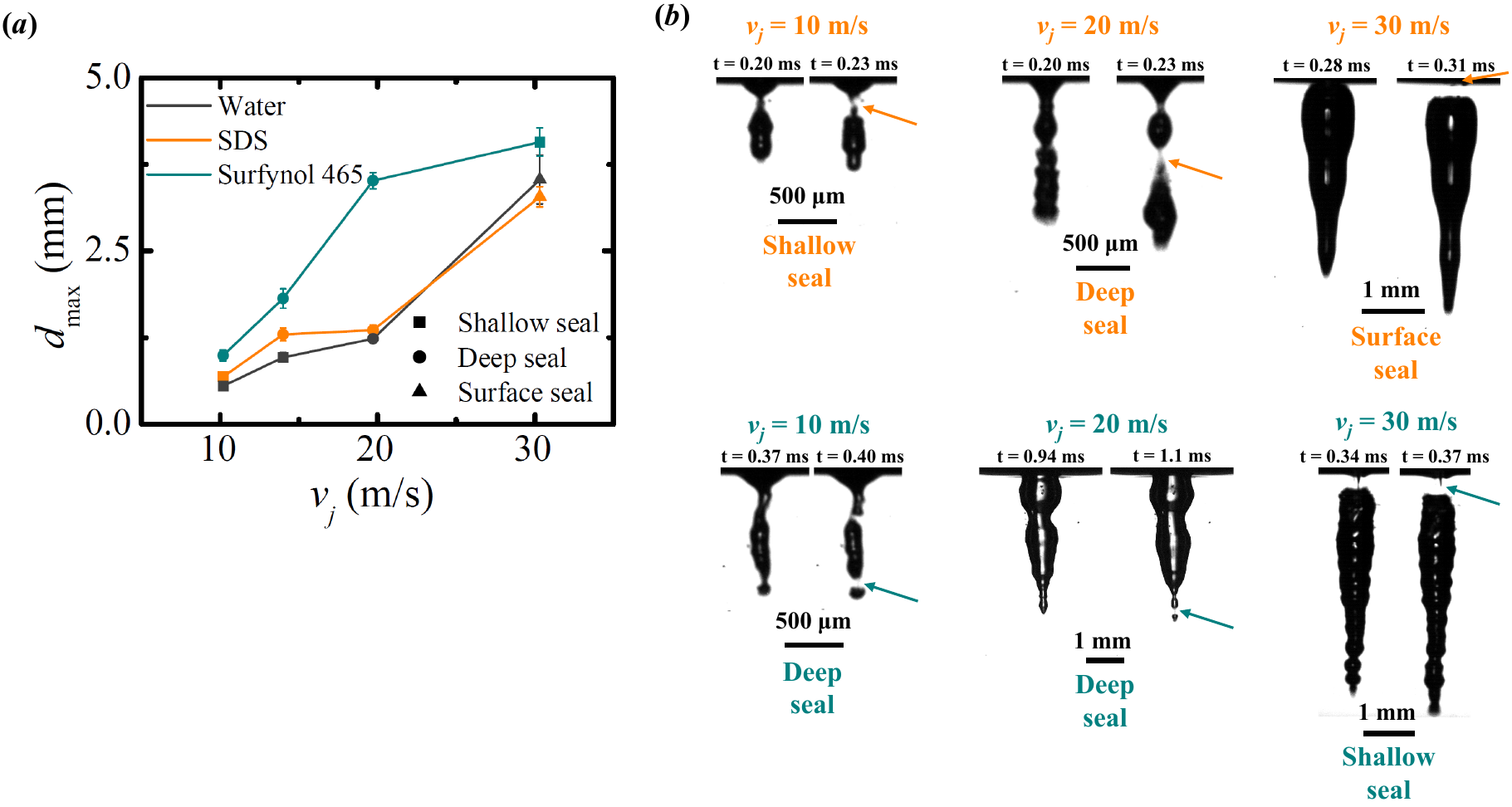}}
\end{center}
\caption{(a) Maximum cavity depth before pinch-off as a function of the jet velocity. (b) Images at the moment of pinch-off for SDS (top) and Surfynol 465 (bottom). The arrows indicate the pinch-off location for each case.}
\label{Max_depth}
\end{figure*}

To further investigate how surfactant kinetics affect the cavity penetration dynamics, we measured the maximum cavity depth reached before the interface pinches over a range of jet velocities, from approximately 10 to 30 m/s. The results are shown in Fig.\ \ref{Max_depth}a. Similar values of the cavity maximum depth were measured for pure water and the aqueous solution of SDS across the investigated velocity range, reinforcing the idea that SDS does not significantly change the effective surface tension during the rapid interface formation. 

For $v_j \approx 10$ m/s, the maximum cavity depth for Surfynol is $0.99\pm 0.08$ mm, nearly twice that in pure water ($0.55\pm 0.04$ mm) and notably greater than that for SDS ($0.68\pm 0.06$ mm). This difference becomes even more apparent at intermediate velocities; for $v_j\approx 20$ m/s, Surfynol allows a cavity depth of $3.52 \pm 0.12$ mm, while the depths in pure water and with SDS are $1.23\pm 0.06$ mm and $1.36\pm 0.08$ mm, respectively (see Fig.\ \ref{Max_depth}a). 

\begin{figure*}[!]
\begin{center}
\resizebox{0.75\textwidth}{!}{\includegraphics{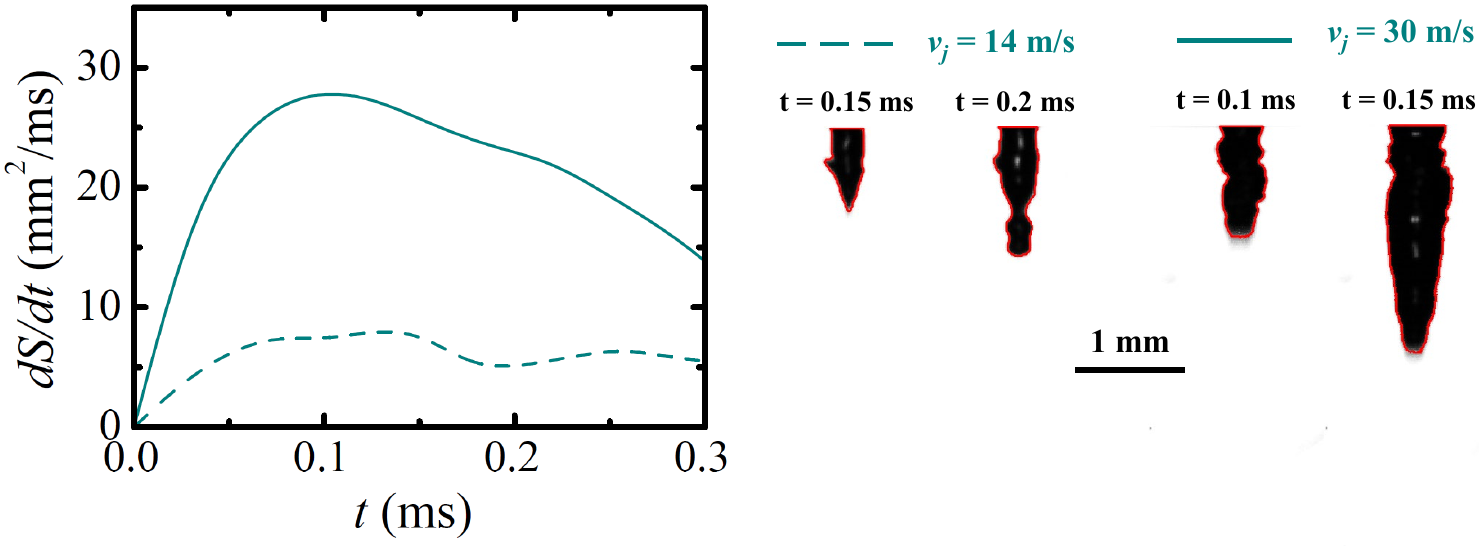}}
\end{center}
\caption{Interface growth rate $dS/dt$ during the cavity evolution in the presence of Surfynol for $v_j=14$ m/s (dashed line) and 30 m/s (solid line), and representative images for the two cases. The red line represents the detected contour at each instant.}
\label{rate}
\end{figure*}

As the jet speed is increased further ($v_j \approx 30$ m/s), the difference between Surfynol and SDS aqueous solutions reduces. Surfynol yields a maximum penetration of $4.07\pm 0.21$ mm, which is similar to the values $3.53\pm 0.35$ mm and $3.28 \pm 0.15$ mm for water and SDS, respectively. For this jet velocity, the cavity displays a prompt surface seal event in the presence of SDS. However, Surfynol produces a tiny air ligament before the surface seal closure, which leads to the shallow seal mode. The air ligament presumably survives due to the Surfynol adsorption at this late stage, which delays the interface pinchoff \citep{KWTB18,PRHEM20}. The similarity between the cavity depths with Surfynol and SDS at high jet velocities indicates a limiting regime, where the new interface is created so quickly that even ultrafast surfactants cannot lower the surface tension within the critical stage of cavity expansion. To illustrate this idea, Fig.\ \ref{rate} compares the interface growth rate $dS/dt$ ($S$ is the interface area) in the presence of Surfynol for $v_j=14$ m/s and 30 m/s. The maximum value of the interface growth rate during the cavity formation for $v_j=30$ m/s is more than three times the value corresponding to $v_j=14$ m/s.

In principle, surfactant adsorption at a new interface can be limited by either diffusion, adsorption, or a combination of both \citep{MS20,M24}. However, the negligible effect of SDS on the cavity dynamics cannot be attributed to limited diffusion. As mentioned in Sec.\ \ref{formulation}, the diffusion velocity for SDS is expected to be similar to that of Surfynol. Therefore, the differences between the two surfactants should be attributed to their adsorption rates. Diffusion can still be a limiting factor in this phenomenon. However, its effect on the cavity dynamics is expected to be essentially the same for the two surfactants.

\subsection{Model for cavity retraction}
\label{sec43}

To quantitatively describe the cavity retraction, we employ a simple nondimensional harmonic oscillator model. Let $\xi=d/d_{\textin{max}}$ and $\tau=t/t_{ic}$ be the normalized cavity penetration depth and dimensionless time, respectively, where $t_{ic}=\sqrt{\rho_b d_{\textin{max}}^3/\sigma_{\textin{ef}}}$ is the inertio-capillary time ($\rho_b d_{\textin{max}}^3$ is proportional to the liquid displaced by the cavity). We assume that the effective surface tension $\sigma_{\textin{ef}}$ is the clean interface value $\sigma=0.072$ N/m for SDS and the equilibrium value $\sigma=0.031$ N/m for Surfynol. We suppose that the penetration depth evolves according to the harmonic oscillator equation
\begin{equation}
\label{eq}
\frac{d^2\xi}{d\tau^2}+2\zeta\frac{d\xi}{d\tau}+\xi=0,
\end{equation}
where $\zeta$ is the dimensionless damping ratio. In this approximation, the cavity retracts due to the restoring force associated with an effective surface tension, and viscosity opposes the retraction. The fitting parameter $\zeta$ plays the role of the effective Ohnesorge number defined in terms of the liquid viscosity, effective liquid mass, and effective surface tension.

Our experiments show that $\xi(\tau)$ corresponds to an overdamped oscillation. In this case, $\zeta>1$, and the solution of Eq.\ (\ref{eq}) is the exponential relaxation function
\begin{equation}
\label{sol}
\xi(\tau) = A \exp(-\lambda_1 \tau) + B \exp(-\lambda_2 \tau),
\end{equation}
where $\lambda_{1,2}=\zeta\pm \sqrt{\zeta^2 -1}$, and $A$ and $B$ are constants determined by the initial conditions.

\begin{figure*}[!]
\begin{center}
\resizebox{1\textwidth}{!}{\includegraphics{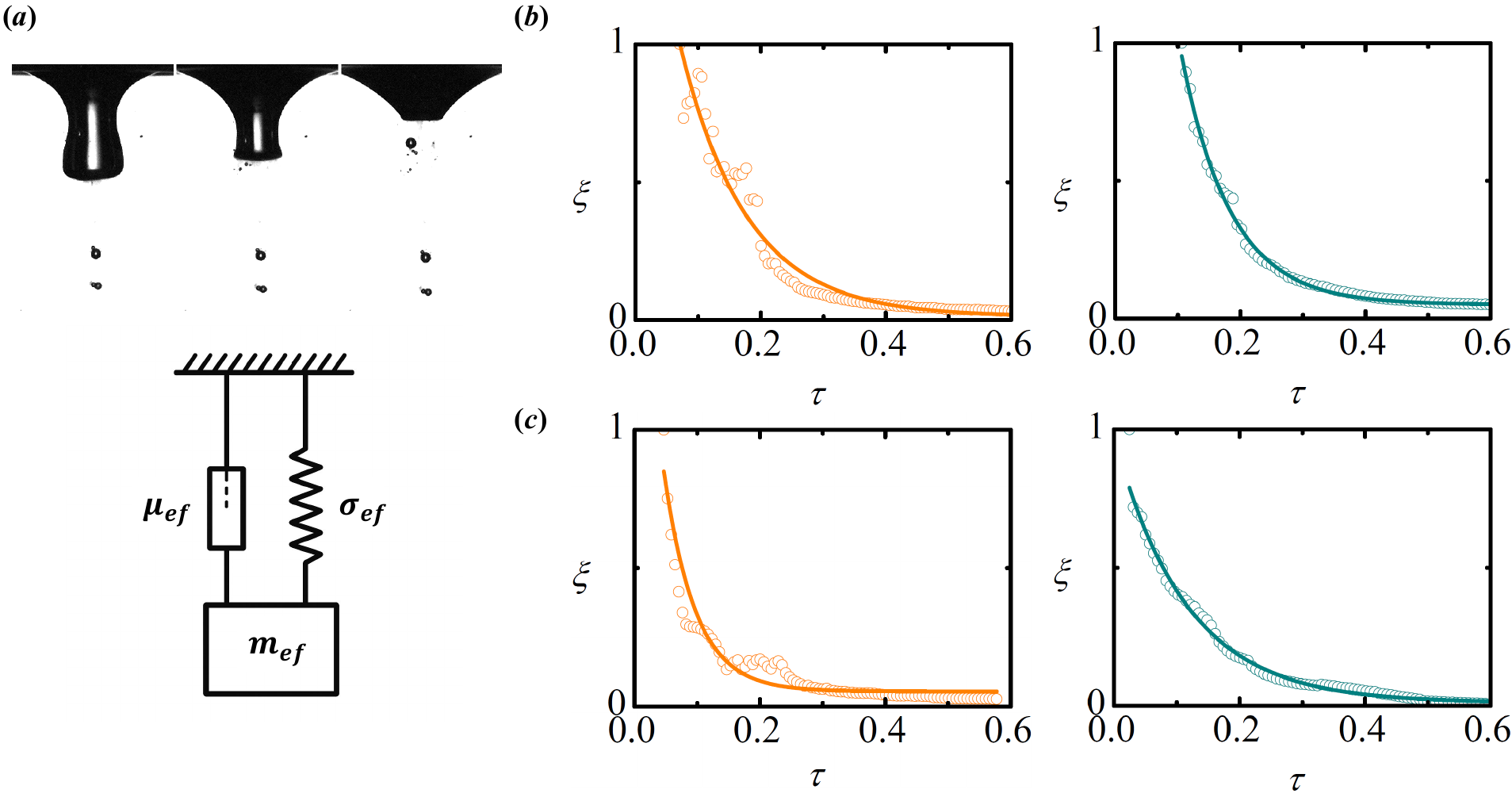}}
\end{center}
\caption{(a) Analogy between a cavity retraction process and a harmonic oscillator. Fit of the model solution to the experimental data for (b) $r_j=32$ $\mu$m and $v_j\approx 14$ m/s, and (c) $r_j=39$ $\mu$m and $v_j\approx 20$ m/s. The left-hand and right-hand graphs correspond to SDS and Surfynol, respectively.}
\label{Fit}
\end{figure*}

In our analysis, we compare the exponential decays in the presence of SDS and Surfynol. We restrict our analysis to intermediate jet velocities for which the same (deep seal) closure mechanism is observed. Table \ref{tab2} shows the values of the only fitting parameter $\zeta$. Figure \ref{Fit} shows the good agreement between the experimental data and the solution (\ref{sol}) to the model (\ref{eq}). 

As mentioned above, the damping ratio $\zeta$ plays the role of an effective Ohnesorge number Oh$_{\textin{ef}}=\mu_b/(m_{\textin{ef}}\, \sigma_{\textin{ef}})^{1/2}$, where $m_{\textin{ef}}=\rho_bd_{\textin{max}}^3$ is the effective value of the mass liquid displaced by the cavity and surface tension. Therefore, the effective surface tension can be calculated as $\sigma_{\textin{ef}}=(\mu_b/\rho_b)\,  \text{Oh}_{\textin{ef}}^{-2} d_{\textin{max}}^{-3}$. Assuming that $\zeta\propto \text{Oh}_{\textin{ef}}$, we conclude that $\sigma_{\textin{ef}}\propto \zeta^{-2} d_{\textin{max}}^{-3}$. Table \ref{tab2} shows the values of $\zeta^{-2} d_{\textin{max}}^{-3}$. As expected, the ratios $(\zeta^{-2} d_{\text{max}}^{-3})_{\textin{Surf}}/(\zeta^{-2} d_{\text{max}}^{-3})_{\textin{SDS}}\approx 0.40$ and 0.35 for $v_j=14$ m/s and 20 m/s, respectively, are close to the ratio $(\sigma_{\text{ef}})_{\textin{Surf}}/(\sigma_{\text{ef}})_{\textin{SDS}} = 31/72 \approx 0.43$ of the efective surface tensions. This agreement confirms our hypothesis that the rapid adsorption of Surfynol plays a key role in reducing the dynamic surface tension.

\begin{table}
    \centering
    \begin{tabular}{ccccccc}
    \hline
    $v_j$ & Surfactant & $\zeta$ & [95\% CI] & $\zeta^{-2} d_{\textin{max}}^{-3}$ \\
    \hline
    14 m/s & SDS      & 4.72 & [4.29, 5.15] & 0.0346 \\
    14 m/s & Surfynol & 6.33 & [6.19, 6.46] & 0.0137 \\
    \hline
    20 m/s & SDS      & 10.05 & [8.59, 11.5]  & 0.00758 \\
    20 m/s & Surfynol & 4.24 & [4.18, 4.31] & 0.00265 \\
    \hline
    \end{tabular}
    \caption{Values of the fitted parameter $\zeta$, the fitting confidence interval, and the value of $\zeta^{-2} d_{\textin{max}}^{-3}$.}
    \label{tab2}
\end{table}
  
\section{Conclusions}
\label{conclusions}

We experimentally investigated the impact of a submillimeter jet on a liquid pool, both with and without surfactants, to explore their effects on this phenomenon. We changed the characteristic time of the cavity formation and collapse by varying the jet speed. Our results show that the maximum cavity depth and timescale of its retraction depend significantly on the rate of surfactant adsorption. As hypothesized, a standard surfactant such as SDS does not significantly affect the cavity dynamics. Conversely, ultrafast surfactants such as Surfynol allow the emergence of deeper cavities that persist longer in the liquid pool. This reinforces the idea that only surfactants with remarkably rapid activity can noticeably alter interface dynamics when a new surface is generated very quickly. In contrast, conventional surfactants typically produce effects only at later stages of this process. Our analysis reveals that Surfynol effects persist across a wide range of impact conditions but ultimately fade at the highest jet speeds. This reveals a kinetic limit where even rapid surfactant action is outpaced by the creation of the interface for the ultrafast surfactant used in this work.

By fitting cavity profiles to the overdamped oscillator model for cases where deep seal closure occurs, we link the mechanical response to an effective dynamic surface tension active during the retraction event. The agreement between the fitted damping ratios and surface tension measurements validates this approach and highlights the importance of rapid adsorption in reducing interfacial tension. 

The capacity of surfactants to dynamically reduce surface tension during rapid interface formation is crucial for controlling fast interfacial flows in both natural processes and technological applications. Here, we have utilized submillimeter jet impacts on surfactant-laden pools as a test bed to investigate the extent to which surfactant adsorption kinetics can influence cavity evolution on submillisecond timescales. However, our conclusions can be applied to a wide range of interfacial phenomena \citep{KRMCAS18, WZL25, VMQF23}. Our findings can inform the design of experiments, revealing the critical role played by fast sorption kinetics. 

\vspace{1cm}

\newpage

\section*{Acknowledgement}
This work was financially supported by the Spanish Ministry of Science, Innovation and Universities (grant no. PID2022-140951OB/AEI/10.13039/501100011033/FEDER, UE).  DF-M acknowledges grant PREP2022-000205 funded by MICIU/AEI /10.13039/501100011033 and ESF+. DFR and UJG acknowledge funding from the project Needle-free injections with file number 19657 of the research programme NWO Talent Programme Vidi TTW, which is financed by the Dutch Research Council (NWO) and funding from the European Research Council (ERC) under the European Union’s Horizon 2020 Research and Innovation Programme (Grant No. 851630).


%

\end{document}